# Composition Anisotropy Drives Large Bulk Photovoltaic Fields Along the Non-polar Vertical Direction in 2D Hybrid Perovskite Ferroelectrics

Yuzhong Hu[1, 2, 3]*, Andrii Shcherbakov[2], Jonathan Zerhoch[2], Lars Schneider[2], Shangpu Liu[2], Xitao Liu[4], Fangping Zhuo[5], Lovro Fulanovic[6], Felix Deschler[2] and Martijn Kemerink[1]*

[1]Institute for Molecular Systems Engineering and Advanced Materials, Heidelberg University, D-69120 Heidelberg, Germany

[2]Physikalisch-Chemisches Institut, Universität Heidelberg, Im Neuenheimer Feld 229, 69120 Heidelberg, Germany

[3]Department of Materials Science and Engineering, NTNU Norwegian University of Science and Technology, NO-7491 Trondheim, Norway

[4]State Key Laboratory of Functional Crystals and Devices, Fujian Institute of Research on the Structure of Matter, Chinese Academy of Sciences, Fuzhou, 350002 Fujian, P.R. China

[5]Department of Materials and Earth Sciences, Technical University of Darmstadt, 64287 Darmstadt, Germany

[6]Institute for Chemistry and Technology of Materials, Graz University of Technology, Stremayrgasse 9, 8010 Graz, Austria

*Corresponding author. Email: huyuzhong1@gmail.com (Y. H.); martijn.kemerink@uni-heidelberg.de (M.K.)

**Abstract**: The photovoltaic electric field of the bulk photovoltaic effect (BPE) reflects the intrinsic ability of ferroelectrics to separate photoexcited excitons into electrons and holes, and are essential parameters for applications such as voltage-readout photodetectors. Because polarization defines the cation-anion displacement and noncentrosymmetric axis, the polar direction is generally one of the orientations exhibiting strong bulk photovoltaic field ($E_{\mathrm{BPE}}$). Here, we report emergent BPE behavior in 2D hybrid perovskite ferroelectrics (HPFs), where $E_{\mathrm{BPE}}$ can be two orders of magnitude higher along the vertical nonpolar direction than along the polar in-plane direction. Its magnitude is up to orders of magnitude higher than that of benchmark photoferroelectrics across different material systems and is the highest among 2D HPFs reported so far. This strong BPE response with emergent directional anisotropy originates from the unique coupling among the shift-current BPE mechanism, an efficient photocarrier-generating inorganic part, and an insulator-like organic part, a combination that is conflicting or inaccessible in traditional photoferroelectrics. This composition and anisotropy also produce basic BPE behavior distinct from that of typical photoferroelectrics, including a laser intensity dependent photovoltage and a nonlinear scaling of photovoltage with material dimension. We analyze and develop a series of formulas to describe the emergent photovoltage phenomena, which should be applicable to this novel 2D ferroelectrics family and to polar systems with similar anisotropy and robust photoelectric response.

**Introduction**: Optical–electrical energy interactions underpin modern photoelectric technologies, and reconfigurable photoelectric states further expand their capabilities toward memory and logic functions[1, 2]. The switchable polarization of ferroelectrics can drive directional separation of photoexcited electrons and holes, producing a reversible steady photocurrent at zero bias, known as the bulk photovoltaic effect (BPE). This makes ferroelectrics promising for memory photoelectronics and can generate BPE photovoltage ($V_{BPE}$) far above the material bandgap, fundamentally distinct from the general photovoltaic effect in pn-junctions[3, 4]. This $V_{BPE}$ and associated BPE field ($E_{BPE}$=$V_{BPE}$/sample dimension) are two key parameters of BPE materials, because they reflect the intrinsic capability of photoferroelectrics to produce an asymmetric potential for separating electrons and holes[5]. $E_{BPE}$ is therefore a primary figure of merit for BPE applications such as photodetectors based on photovoltage. Hence, identifying photoferroelectrics capable of sustaining large $E_{BPE}$ is central to advancing BPE-based optoelectronics[6]. BPE parameters such as $E_{BPE}$ are raised by either ballistic-current or shift-current mechanisms, depending on the system dimension. At bulk length scales above the micrometer regime, the ballistic contribution vanishes, and BPE is governed by the shift-current mechanism[7, 8]. Specifically, because of the asymmetric potential in ferroelectrics, light-induced carrier excitation shifts the average position of the electronic wave packet toward the lower-potential side, producing a real-space displacement known as shift current[7, 9]. Because polarization reflects the relative displacement of cations and anions and defines the noncentrosymmetric axis, the polar direction is generally regarded as one of the orientations with the strongest BPE response in traditional single-domain photoferroelectrics[10-12]. A BPE response along nonpolar or other directions far exceeding that along the polar direction is therefore rarely observed[10, 13].

Here, we report emergent BPE behavior in $EA_4Pb_3Br_{10}$ (EPB, EA = ethylammonium), a 2D hybrid perovskite ferroelectric (HPF) with in-plane polarization. The combination of BPE shift current mechanism, organic-inorganic composition and anisotropy gives this 2D HPF abundant photoexcited carriers, low out-of-plane conductivity, and a highly nonuniform illumination profile, which are self-conflicting in traditional ferroelectrics but ideal for generating a large $E_{BPE}$ and lead to a photovoltage response that is distinct from typical BPE behavior. We find that $E_{BPE}$ along the nonpolar out-of-plane direction (238 V/cm) is two orders of magnitude higher than that along the polar in-plane direction. To the best of our knowledge, such behavior has not been reported in previous photoferroelectric materials. This $E_{BPE}$ is also the highest among reported 2D HPFs with > 50 compositions designed by ferroelectrochemistry methods and experimentally demonstrated[14]. It is also several times larger than that of benchmark traditional photoferroelectrics across different material systems, including $BiFeO_3$[15], SbSI[16] and tetrathiafulvalene-p-chloranil (TTF-CA)[10]. In addition, this 2D HPF exhibits basic photoferroelectric responses that are strikingly different from general BPE behavior, including laser intensity dependent (rather than independent) photovoltage and a nonlinear (rather than linear) relationship between photovoltage and sample dimension. The present study advances photoferroelectric research by enhancing an essential photoferroelectric parameter ($E_{BPE}$) through a mechanism that is distinct from previous reports, identifying emergent basic BPE behavior, and providing a formula framework for this novel 2D photoferroelectrics as well as polar systems with similar anisotropic and robust photoelectric features. For HPFs, this work reveals their promise for voltage-readout photoferroelectric devices and should help explain the recently discovered large field effect of 2D HPFs on contiguous 2D materials along the nonpolar vertical direction under illumination[17].

The single crystal of EPB (Supplementary Fig 1a)[18] was synthesized by a temperature-lowering solution method similar to a previous report[19] with the composition confirmed by X-ray diffraction (see Supplementary Fig 1b) [18]. The structure and in-plane polarization configuration are shown in Fig. 1a. The 3D PbBr inorganic frameworks are separated by EA organic components, with van der Waals bonds linking adjacent unit layers. Spontaneous polarization can be identified by the long-range ordering of organic cations and inorganic anions. Similar to most 2D HPFs, the polarization projections along vertical direction cancel each other, leaving a net polarization along the in-plane direction (see Supplementary Fig 2[18] for detailed phase transition and biaxial polarization identification)[20]. The ferroelectricity of the crystal is confirmed by ferroelectric hysteresis-loop measurements along the polar direction at room temperature. A coercive field around 20 kV/cm and a remnant polarization of around 4 μC/cm$^2$ are identified from the polarization-electric field (*PE*) loop, in good agreement with a previous report[19].

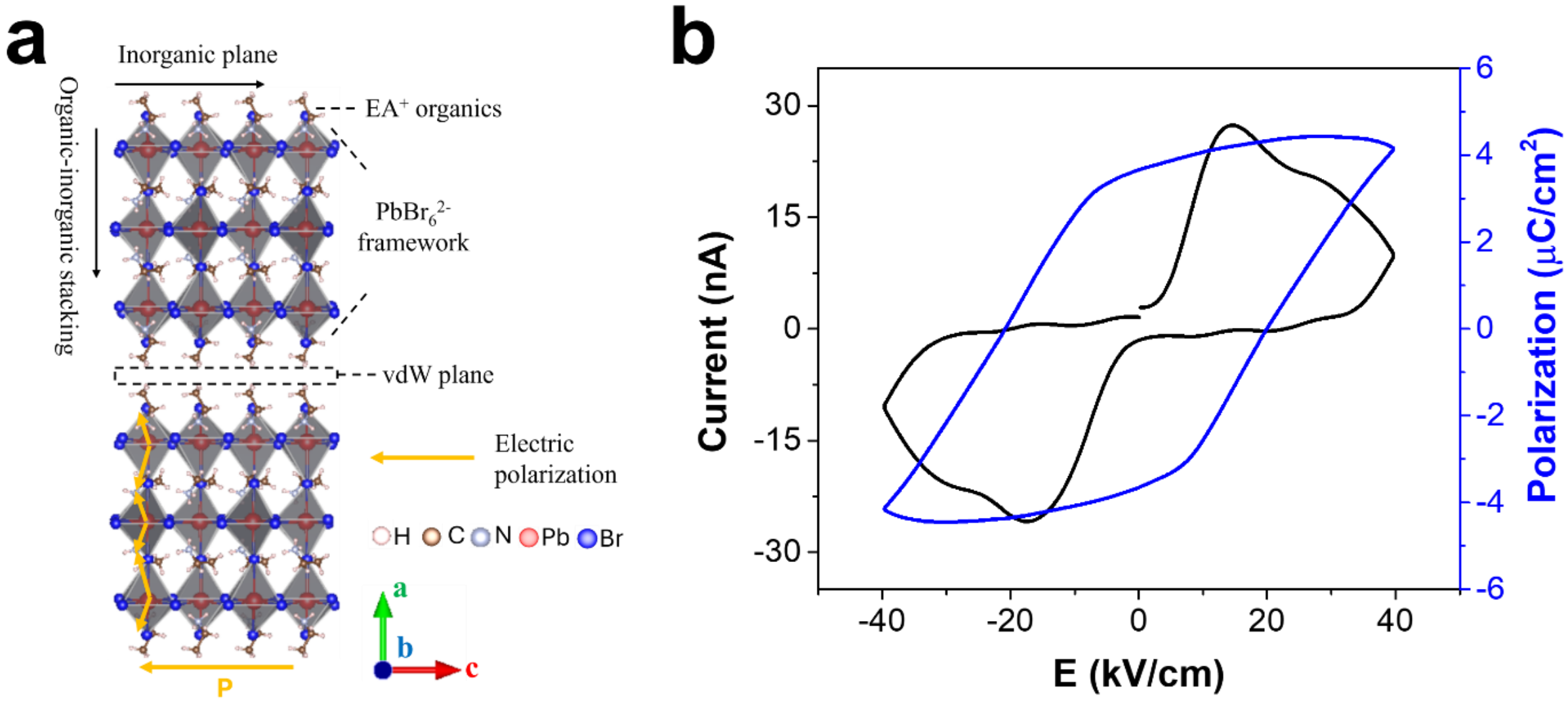

**Fig.1. Crystal structure and ferroelectric properties of $EA_4Pb_3Br_{10}$. (EPB). a**, Crystallographic structure viewed along the *c*-axis, showing the layered Ruddlesden–Popper–type hybrid perovskite with composition and structural anisotropy. Similar to other HPFs, the $N^+$ site in the organic layer and the centroid of the $PbBr_6^{2-}$ octahedron can be taken as the cation and anion centers[21]. The corresponding electric dipole, indicated by orange arrows, gives rise to in-plane spontaneous polarization. Crystallographic data are from a previous report[19]. **b**, Ferroelectric hysteresis loop and the corresponding switching current–electric field loop measured along the in-plane polar direction of the bulk crystal at room temperature with frequency of 4 Hz.

BPE was first investigated along the in-plane polar direction under 405 nm linear polarized laser with the sample and laser geometry shown in Fig. 2a. In this coordinate system and for the $C_{2v}$ point group of EPB, the BPE photocurrent can be described as follows (see details in Supplementary Note) [18, 22]:

$$J_c = \frac{1}{4}E^2\{\cos 2\varphi\,[\beta_{caa}(1-\cos 2\theta) - 2\beta_{cbb} + \beta_{ccc}(\cos 2\theta + 1)] + \cos 2\theta(\beta_{ccc} - \beta_{caa}) + \beta_{caa} + 2\beta_{cbb} + \beta_{ccc}\} \quad (1)$$

Where $E$, $\varphi$, $\beta_{ijk}$, and $\theta$ are the electric-field amplitude of the laser, the laser polarization angle, the BPE tensor, and the incident angle with respect to the surface normal, respectively. This indicates that, similar to general BPE behavior, $J_c$ contains an offset term and shows a clear $\cos 2\varphi$ dependence with a period of 180 degrees, which is confirmed by the polarization-angle-dependent $J_{BPE}$ measurement (Fig. 2b).

Another fingerprint of BPE is a switchable photocurrent. As shown in Fig. 2c, a clear ferroelectric-polarization dependence is observed when the crystal is poled in opposite directions. A $V_{BPE}$ = 0.2 V is identified, similar to other 2D HPFs[23, 24]. Noticeably, the $J_{BPE}$ = 0.25 mA/cm$^2$ under laser intensity $I_0$ = 6.4 mW/cm$^2$ corresponds to a BPE coefficient $\beta^L$ ($J_{BPE}/I_0$) of 39 μA/mW, which is much higher than those of high-performing traditional photoferroelectrics such as $BiFeO_3$ and $LuMnO_3$ for which values in the range $10^{-2}$ - $10^{-1}$ μA/mW [25-27]. This reflects the high-performance photoelectric character of the inorganic lead-halide component in 2D HPFs. In the above, the effective electrode area is estimated from the electrode length multiplied by the light penetration depth characterized using a spin-coated thin film (see Supplementary Fig. 3 for details) [18]. The laser-power-dependent BPE response was then characterized, c.f. Fig. 2d, with the same structure, showing a typical $I_0$-proportional $J_{BPE}$ and an $I_0$-independent $V_{BPE}$[28].

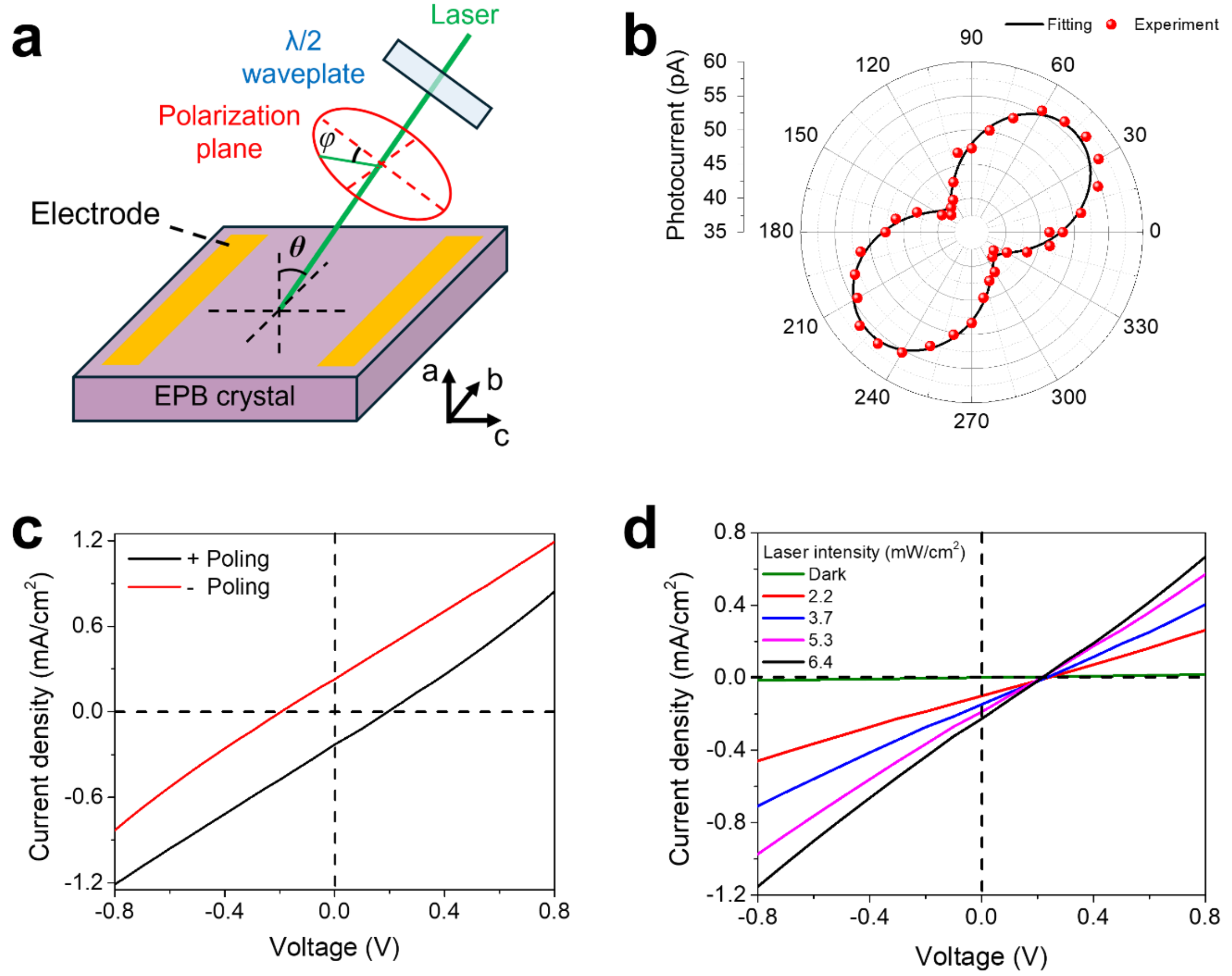

**Fig. 2. Bulk photovoltaic characterization of EPB along the 2D in-plane polar direction under linearly polarized 405 nm laser illumination. a**, Coordinate system and experimental setup for the measurements. The linearly polarized laser lies in the *ac* plane with an incident angle ($\theta$) of 30˚. **b**, Bulk photovoltaic photocurrent at different light polarization angles ($\varphi$). The line indicates a fit to the $I_{BPE}$ magnitude= $a_1 + b_1 \times sin2\varphi$. **c**, Ferroelectric-polarization-dependent $J_{BPE}$ -$V$ behavior. **d**, Laser intensity dependent $J_{BPE}$ -$V$ behavior.

Next, the BPE behavior was investigated along the out-of-plane nonpolar direction using a top-bottom structure with 3.5 nm thin Au electrodes for high transparency. Here, the BPE photocurrent can be described as follows (see details in Supplementary Note) [18]:

$$J_a = \frac{1}{4}\beta_{aac}E^2 \sin 2\theta(1 + \cos 2\varphi) \quad (2)$$

Similar to the in-plane BPE investigation, we performed $\varphi$-dependent photocurrent measurements, which show a clear 180-degree period (Fig. 3a). Unlike equation (1), in which $J_a$ contains relatively complex terms, equation (2) indicates that $J_a$ is directly proportional to the incident angle $\theta$ as $J_a \propto \sin 2\theta$, as confirmed by our $\theta$-dependent BPE measurement (Supplementary Fig. 4a-c) [18]. Noticeably, the nearly vanished photocurrent at $\theta$ = 0 indicates a zero-offset photocurrent, consistent with equation (2). This differs from the relatively large offset term ($J_{c\text{-offset}} = \frac{E^2}{4}[\cos 2\theta(\beta_{ccc} - \beta_{caa}) + \beta_{caa} + \beta_{cbb} + \beta_{ccc}]$) in equation (1) and from the experimental result in Fig. 2b ($J_{offset}$ = 48.14 pA). The BPE response was also characterized at $\theta = \pm 30^\circ$, where the direction of $J_{BPE}$ shows a direct reversal (Fig. 3b). Surprisingly, the photovoltage obtained along this nonpolar direction (5.4V) is much larger than that along the polar in-plane direction. Its magnitude, far above the bandgap of EPB (2.6 eV), further confirms its BPE origin.

We next investigate the laser intensity dependent BPE response. Unlike typical BPE behavior, the out-of-plane $V_{BPE}$ shows an almost linear, rather than independent, relationship to $I_0$ (Fig. 3c). This emergent behavior weakens and eventually almost vanishes as the thickness decreases (Fig. 3d). In addition, the $V_{BPE}$- sample thickness ($d$) relationship is clearly nonlinear (Fig. 3e), in contrast to the linear relation ($V_{BPE} = E_{BPE} \times d$) in typical BPE behavior. Nevertheless, the obtained $E_{BPE}$ along this vertical nonpolar direction is remarkably large (238 V/cm). This value is two orders of magnitude larger than that of EPB along the polar direction (8.8V/cm) and all 2D HPFs reported so far, and several times higher than those of high-performance photoferroelectrics such as $BiFeO_3$ and $(4x\text{-}BA)_2(EA)_2Pb_3Br_{10}$. The former is one of the most intensively investigated traditional photoferroelectrics and the latter are the HPF with the highest reported $V_{\mathrm{BPE}}$[29], as summarized in Fig. 3f.

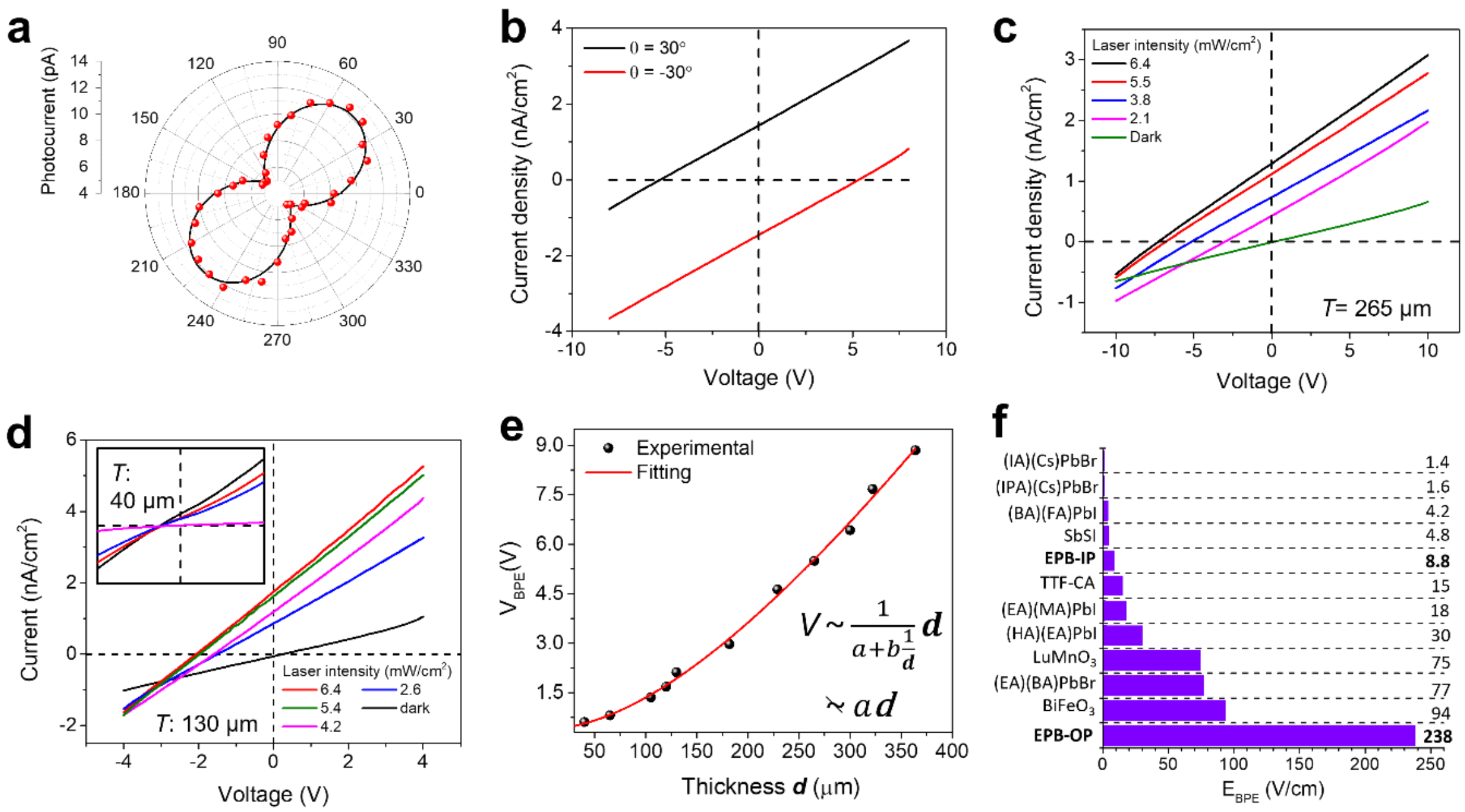

**Fig. 3. Bulk photovoltaic characterization along the 2D out-of-plane (nonpolar) direction. a**, Bulk photovoltaic photocurrent at different light polarization angles. The line indicates a fit to the $I_{\mathrm{BPE}}$ magnitude $= a_2 + b_2 \times sin2\varphi$. **b**, $J_{\mathrm{BPE}}$ measured at $\theta = \pm\, 30^{\circ}$. **c**, **d**, Laser intensity dependent $J_{\mathrm{BPE}}$-$V$ behavior of EPB crystals with thicknesses of 265, 130, and 40 μm (inset). Details of the inset are shown in Supplementary Fig. 4d[18]. **e**, $V_{\mathrm{BPE}}$-thickness relationship. The line indicates a fit of $V_{\mathrm{BPE}}$ magnitude $= c_1 + d/(a_3 + b_3 \times \frac{1}{d})$, c.f. Eq. 12. **f**, $E_{\mathrm{BPE}}$ comparison of HPFs and classic photoferroelectrics taken from related works Refs. [10, 15, 16, 23, 24, 27, 29-32].

We now analyze these emergent phenomena and develop formulas to describe their unconventional BPE behavior. As discussed in the Supplementary Note[18], $J_l = \sum_{ij} \beta_{ijl}\, (E_a E_b^* + E_a^* E_b)/2$. This can also be expressed as

$$J_{BPE} = \beta^L I_0 \quad (3)$$

where $\beta^L$ and $I_0$ are the linear BPE coefficient and the laser intensity[7]. Photoconductivity can be expressed as $\sigma_{phe} = n\mu e = eI\phi\mu\tau/h\nu$, where $n$, $e$, $\phi$, $\mu$, $\tau$, $h$, $\nu$ are carrier density, elementary charge, quantum yield, carrier mobility, carrier lifetime, Planck's constant and frequency of light, respectively[33]. In general, the $V_{BPE}$ of traditional photoferroelectrics can be described by

$$V_{BPE} = \frac{dJ_{BPE}}{\sigma_{phe}} = \frac{\beta^L d\hbar\omega}{e\phi\mu\tau}\frac{I_0}{I_0} = \frac{\beta^L dh\nu}{\phi\mu\tau} \quad (4)$$

$V_{BPE}$ is therefore independent of $I_0$, with typical behavior as shown in Fig. 2b. However, this assumption requires the overall conductivity, $\sigma = \sigma_{phe} + \sigma_{dark}$, to be dominated by $\sigma_{phe}$, so that $\sigma \approx \sigma_{phe}$. This condition is fulfilled in most previous photoferroelectric. For traditional photoferroelectrics with poor semiconducting properties such as Fe-doped $LiNbO_3$, a low dark conductivity (~$10^{-9}$ S/m) is generally accompanied by a low absorption coefficient (~$10^1$/cm) in the visible range[34, 35]. This corresponds to a light penetration depth of around 1 mm, indicating that photoconductivity, rather than dark conductivity ($\sigma_{dark}$), governs the material response for typically used sample dimensions. For photoferroelectrics with good semiconducting properties, such as $BiFeO_3$, a high absorption coefficient (~$10^4$/cm) is generally coupled with a higher conductivity (~$10^{-7}$ S/m), or the response is mostly studied in in-plane thin-film structures, where BPE is weakly affected by the nonilluminated region and $\sigma_{dark}$[36-38].

However, as shown in Fig. 4a, when measured along the out-of-plane direction, 2D HPF become a special combination of an insulator-like organic part and an efficient photocarrier-generating lead-halide part, a compositional feature impossible for traditional photoferroelectrics. This produces apparently contradictory yet integrated properties: a high absorption coefficient of 1.06 × $10^5$/cm (Supplementary Fig. 3[18], corresponding light penetration depth of 94 nm), a high photoelectric conversion efficiency, reflected by the large $\beta_L$ and $J_{BPE}$ in the in-plane measurement, together with an out-of-plane conductivity that is much lower than that of oxide ferroelectrics including $LiNbO_3$ ($1.59 \times 10^{-10}$ S/m from the dark current in Fig. 3c). The former two are features of good ferroelectric semiconductors, whereas the latter is characteristic of poor ferroelectric semiconductors. This unconventional combination yields a highly nonuniformly illuminated system with a large density of carriers generated by the shift current mechanism in a thin surface layer, connected in series with a bulk region that is dominated by a very low $\sigma_{dark}$. Such a system is ideal for producing large $E_{BPE}$ and is strikingly different from general photoferroelectrics, giving rise to emergent $V_{BPE}$ behavior.

Here, we explain the giant and anisotropic $E_{BPE}$ as well as emergent thickness dependent $V_{BPE}$ through a formula analysis coupled with the experimental results. As discussed above, because of the high absorption and a light penetration depth below 100 nm, a bulk crystal thicker than the micrometer scale forms a highly nonuniformly illuminated system. Under this condition, at thickness $x$, the local photoconductivity is governed by the absorption coefficient and residual light intensity $I_x$. From the Beer–Lambert law, $I_x = I_0 \times \exp(-\alpha x)$ , the absorbed light intensity per unit depth is:

$$-\frac{dI}{dx} = \alpha I_0 \exp(-\alpha x) \tag{5}$$

Converting this to the number of absorbed photons using $hv$ and the quantum yield gives the photoexcited carrier generation rate $g$ at $x$:

$$g_x = \phi \frac{\alpha I_0 \exp(-\alpha x)}{hv} \tag{6}$$

With $n = g_x\tau$ ( $g_x$ the local photoexcited carrier generation rate at depth $x$), the local photoconductivity is

$$\sigma_{xphe} = n\mu e = e\mu\tau\frac{\alpha\phi I_0 \exp(-\alpha x)}{h\nu} \tag{7}$$

We define an effective photoconductivity by integrating and averaging over the thickness. This parameter reflects the average $\sigma_{\text{phe}}$ of the crystal, whose different regions possess different photoconductivities, and is used to describe BPE-related parameters and their relationships:

$$\sigma_{phe-eff} = \frac{1}{d}\int_0^d \sigma_{xphe}\, dx = \int_0^d e\mu\tau\frac{\alpha\phi I_0 \exp(-\alpha x)}{dh\nu}\, dx$$

$$= \frac{e\mu\tau\phi I_0}{h\nu d}(1-\exp(-\alpha d)) \tag{8}$$

When the thickness is small, $\alpha d \ll 1$, and $1-\exp(-\alpha d) \approx \alpha d$, thus:

$$\sigma_{phe-thin} = \frac{e\mu\tau\phi\alpha I_0}{h\upsilon} \tag{9}$$

This indicates that with certain $\alpha$ and $I_0$, $\sigma$ shows no dependency on thickness $d$, indicating a uniform illumination condition within the light penetration depth. With high thickness, $1 - \exp(-\alpha d) \approx 1$, thus:

$$\sigma_{phe-thick} = \frac{e\mu\tau\phi I_0}{h\upsilon d} \tag{10}$$

Noticeably, $\sigma$ now shows no dependency on $\alpha$ as the light will always be fully absorbed, giving rise to a highly nonuniform illumination condition. With equations (3), (4), and (5), $V_{\text{BPE}}$ within the penetration depth ($\sigma \approx \sigma_{phe-thin}$) and beyond penetration depth ($\sigma = \sigma_{phe-thin} + \sigma_{dark}$), $V_{BPE} = J_{BPE}/\sigma$ becomes:

$$V_{thin} \approx \frac{\beta^L h\nu d}{\phi e \alpha \mu \tau} \tag{11}$$

$$V_{thick} = \frac{\beta^L I_0}{\sigma_{dark} + \frac{e\mu\tau\phi I_0}{h\nu} \times \frac{1}{d}} d \tag{12}$$

As indicated by equation (10), for samples thicker than the penetration depth (~94 nm), the effective photoconductivity $\sigma_{phe\text{-}eff}$ is proportional to $1/d$. Thus, photoconductivity will decrease sharply with thickness. Combined with the very low electric conductivity of 2D HPFs along the out-of-plane direction, the effective conductivity will then be strongly influenced and ultimately dominated by the low dark electric conductivity, which is reflected by the similar *I*-*V* slopes in the dark and under illumination for thick samples as observed in Fig. 3c. More importantly, this renders a highly advantageous system for producing large $E_{BPE}$ ($J_{BPE}/\sigma$) along the vertical direction. In contrast, when measured along the polar in-plane direction, the large number of photoexcited carriers produced by the shift current mechanism will be transported through the well-conducting inorganic plane, which is characterized by high $\sigma$ and only a moderate $E_{BPE}$ will be produced. This accounts for the large $E_{BPE}$ and strong anisotropy in this 2D HPF.

As discussed, in very thick sample, $\sigma$ along vertical direction will be ultimately dominated by $\sigma_{dark}$, and $V_{BPE}$ becomes:

$$V_{BPE} \approx \frac{J_{BPE}}{\sigma_{dark}} = \frac{\beta^L}{\sigma_{dark}} I_0 \propto I_0 \tag{13}$$

This leads to the nearly linear relationship between $I_0$ and $V_{BPE}$ observed in Fig. 3c. As the crystal becomes thinner, the contribution from photoconductivity increases and eventually dominates $\sigma$. The *J*-*V* curves therefore become increasingly similar to the typical BPE behavior case described by equation (11), where $V_{BPE}$ is independent of $I_0$. This transition is reflected in Fig. 3c-d, where the $J$-$V$ behavior gradually approaches that of traditional photoferroelectrics. In addition, because of this compositional anisotropy and the pronounced influence of $\sigma_{\mathrm{dark}}$, $V_{\mathrm{BPE}}$ does not show the typical linear dependence on sample dimension $d$ described by equation (11). Instead, it follows equation (12), $V_{\mathrm{BPE}} \sim \frac{1}{a+b\frac{1}{d}} \times d$, as observed in Fig. 3e. Equations (11)-(13) and the related analysis explain the large $E_{\mathrm{BPE}}$, its anisotropy, and the emergent $V_{\mathrm{BPE}}$-$I_0$ and $V_{\mathrm{BPE}}$-$d$ behaviors of EPB. They shall be applicable to other 2D HPFs and similar anisotropic polar systems with high photoelectric robustness.

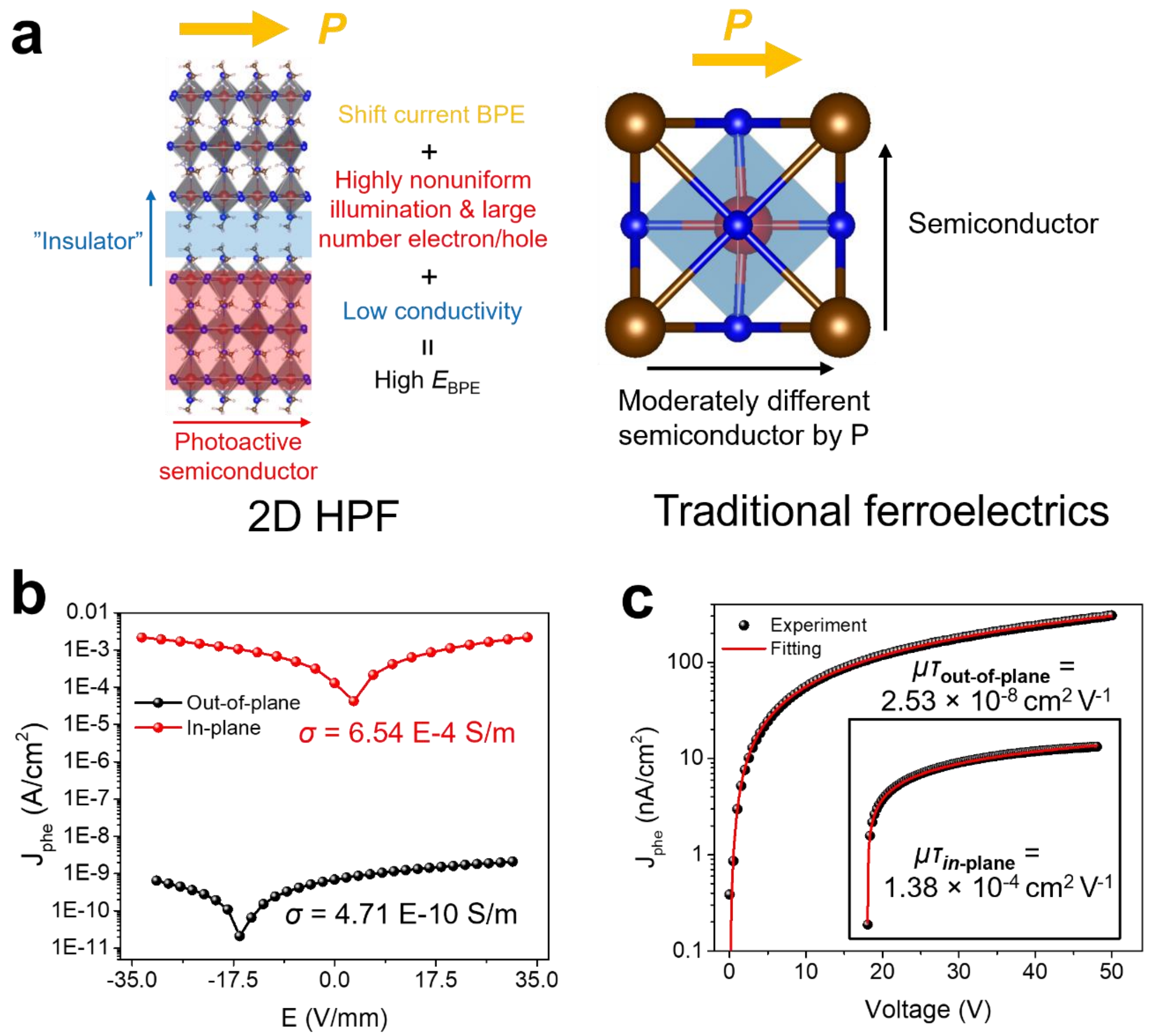

**Fig. 4. The structural, composition, photoelectric properties anisotropy of EPB and comparison with traditional photoferroelectrics.** **a**, Comparison of the anisotropy and composition of EPB and traditional oxide perovskite photoferroelectrics (crystallographic data from a previous report[39]). **b**. $J_{phe}$ – electric field ($E$) curves and photoconductivity along the in-plane and out-of-plane directions, the thickness of out-of-plane sample is 265 μm. **c**. $J_{phe}$ – $V$ curves and extracted $\mu\tau$ along in-plane (inset) and out-of-plane directions. The thickness of the out-of-plane sample is 110 μm. Lines are fit to Eq. 14 and 15. Details of the inset are shown in Supplementary Fig. 5[18].

Finally, we conduct more photoelectric measurements to quantitatively clarify this photoelectric anisotropy. As shown in Fig. 4b, the effective photoconductivities along the in-plane and out-of-plane directions are $6.54 \times 10^{-4}$ and $4.71 \times 10^{-10}$ S/m, respectively. Interestingly, the former is much higher, whereas the latter is far lower than the photoconductivities of typical good (~$10^{-6}$ S/m for $BiFeO_3$) and poor ferroelectric semiconductors (~$10^{-8}$ S/m for Fe-doped $LiNbO_3$) under similar laser intensity[34, 40]. This reflects the competing yet coupled high photoelectric activity of the inorganic component and insulating nature of the organic component. To the best of our knowledge, such a six order of magnitude photoconductivity anisotropy has not been observed in other photoferroelectrics, which generally show $\sigma_{phe}$ differences of only a few to several tens of percent along different directions because of the relatively isotropic 3D oxide perovskite nature (see Fig. 4a)[41]. Photoelectric performance is also characterized by the figure of merit $\mu\tau$ ($\mu$: mobility, $\tau$: charge carrier lifetime), which largely governs $\sigma_{phe}$ and can be extracted from the $J$-$V$ curve by fitting with the Hecht equations for in-plane and out-of-plane structures[42, 43]:

$$J_{in-plane} = \frac{\mu\tau\phi I_0 e}{w^2 dLh\nu}\left(1 + \frac{\mu\tau V}{w^2}\left(\exp\left(-\frac{w^2}{\mu\tau V}\right) - 1\right)\right) \quad (14)$$

$$J_{out-of-plane} = \frac{I_0 \mu\tau V}{Ad^2} \times \frac{1-\exp\left(-\frac{d^2}{\mu\tau V}\right)}{1+\frac{dv_s}{V\mu}} \quad (15)$$

Where $w$, $L$, $A$ and $v_s$ are the sample width, electrode length, device area, and recombination velocity, respectively. As shown in Fig. 4c, the $\mu\tau$ values along the in-plane and out-of-plane directions are 1.38 $\times 10^{-4}$ and 2.53 $\times$ $10^{-8}$ $cm^2V^{-1}$, respectively, consistent with the strongly anisotropic photoconductivity of EPB.

In summary, our work (1) significantly advances bulk photovoltaic research by revealing a large essential photoferroelectric property ($E_{BPE}$) through mechanisms that are distinct from and inaccessible to traditional materials; (2) develops a series of formulas that provide a foundation for understanding the emergent photoferroelectric behavior of 2D HPFs, which differ strikingly from typical BPE behavior and should also apply to similar anisotropic polar systems with strong photoelectric robustness; and (3) reveals the tantalizing potential of burgeoning 2D HPFs family for voltage-readout photoferroelectric applications. Such $V_{BPE}$ much higher than bandgap is not accessible by non-ferroelectric hybrid perovskite. This study should also help explain the recently discovered large field effect along nonpolar vertical direction of 2D HPFs on contiguous 2D materials under illumination[17].

## Acknowledgments

We acknowledge support from the Alexander von Humboldt Foundation. Y.H. thanks Prof. Marin Alexe (University of Warwick) for valuable discussions on the bulk photovoltaic behavior and mechanism, Prof. Jürgen Rödel (Technical University of Darmstadt) for access to the high voltage ferroelectric tester facility.

## Competing Interests Statement

The authors declare no competing interests.